\documentclass[11pt,epsf]{article}
\usepackage{amsmath}
\usepackage{amsfonts}
\usepackage{amssymb}
\usepackage{graphicx}
\usepackage{color}

\newcommand {\dfn} {\stackrel{\Delta} {=}}
\newcommand {\exe} {\stackrel{\cdot} {=}}

\newcommand {\bE} {\mbox{\boldmath $E$}}

\newcommand{\calH}{{\cal H}}

\newcommand{\calP}{{\cal P}}

\begin{document}
\thispagestyle{empty}
\title{Optimal Correlators for Detection and Estimation in Optical Receivers
%\thanks{This research was supported by my wife and kids.}
}
\author{Neri Merhav
%\thanks{
%Currently on sabbatical leave at HP Laboratories,
%1501 Page Mill Road, MS 3U-4, Palo Alto CA 94304, USA.}
}
\date{}
\maketitle

\begin{center}
The Andrew \& Erna Viterbi Faculty of Electrical Engineering\\
Technion - Israel Institute of Technology \\
Technion City, Haifa 32000, ISRAEL \\
E--mail: {\tt merhav@ee.technion.ac.il}\\
\end{center}
\vspace{1.5\baselineskip}
\setlength{\baselineskip}{1.5\baselineskip}

\begin{abstract}
Motivated by modern applications of light detection and ranging (LIDAR),
we study the model of an optical receiver based on an avalanche photo-diode (APD), followed by
electronic circuitry for detection of reflected optical signals and estimation of their delay.
This model is known to be quite complicated as it
consists of at least three different types of noise: thermal noise, shot noise, and multiplicative noise
(excess noise) that stems from the random gain associated with the photo-multiplication of the APD. 
Consequently, the derivation of the optimal likelihood ratio test (LRT) 
associated with signal detection is a non--trivial task, which has no 
apparent exact closed--form solution.
We consider instead a class of relatively simple detectors, that are based on correlating the noisy received signal with
a given deterministic waveform, and our purpose 
is to characterize the optimal waveform in the sense of the best trade--off between the
false--alarm (FA) error exponent and the missed--detection (MD) error
exponent. In the same spirit, we also study the
problem of estimating the delay on the basis of maximizing the correlation between the received signal and
a time--shifted waveform, as a function of this time shift. We characterize the optimal correlator waveform
that minimizes the mean square error (MSE) in the regime of high signal--to--noise ratio (SNR).
The optimal correlator waveforms for detection and for estimation turn out to be different, 
but their limiting behavior is the same: 
when the thermal Gaussian noise is dominant, 
the optimal correlator waveform becomes proportional to the clean signal, but when the thermal noise is 
negligible compared to the other noises, then it becomes logarithmic function of the clean signal, as expected.\\

\end{abstract}

\newpage
\section{Introduction and Background}

The concept of light detection and 
ranging (LIDAR), which can be thought of as the optical analogue of radar, 
is by no means new, and during the many years it has been in use,
it has found an extremely large variety of applications in a wide spectrum of
areas and disciplines, including: agriculture,
archaeology, biology, astronomy, geology and soil sciences, forestry, 
meteorology, and military applications, just to name a few.
Most notably, there are several modern technologies that involve LIDAR, such as
autonomous vehicles, space flight devices, robots of many kinds, 
systems with GRID based processing, and in the future, and face recognition in
biometric systems (e.g., at airports).

This background sets the stage and motivates renewed interest in optical signal detection and estimation.
The customary model of a direct--detection optical receiver (or detector) consists of a 
photo-diode (PIN diode or avalanche photo-diode), that converts the intensity of the received 
optical (laser) signal, modeled as the
rate function of a variable--rate Poisson process, into a train of current impulses 
generated by the photo--electrons at random time instants, pertaining to the Poisson arrivals.
This current is then fed into some electronic circuitry, whose 
first stage is normally a trans-impedance amplifier
(TIA), that amplifies the current signal and converts it into a relatively strong voltage signal, but this
amplification comes at the cost of some distortion as well as thermal noise associated with the 
electronic circuitry. 

The challenge in the development of a solid 
theory of detection and estimation, for such a signal--plus--noise system, is that it is has a
rather complicated model due to the various types of noise involved. 
The combination of shot noise, due to the photo-diode, and the thermal electronic noise is already not trivial.
In the case of an avalanche photo--diode (APD), which is the more relevant case, 
there is an additional, third type of noise, namely, the {\it excess noise}, induced by the APD, 
which is actually a multiplicative noise 
process pertaining to fluctuations in the random gain associated with the
avalanche mechanism, as each primary 
electron--hole pair (generated by a photon absorbed in the photo-diode), 
may generate secondary electron--holes, 
which in turn can generate additional electron--holes, and so on.
For more details, and additional aspects of the problem area, the interested reader is 
referred to some earlier work, e.g., 
\cite{BL90}, \cite{Einarsson96}, \cite{EH81}, \cite{FGS75}, \cite{GJN09},
\cite{GP87}, \cite{MS76}, \cite{Olsson89},
\cite{Personick71a}, \cite{Personick71b}, \cite{Personick08}, \cite{Salz85},
\cite{SSH18}, \cite{TM94}, and \cite{YLC19},
which is by no means an exhaustive list of relevant articles (and a book).

To provide just a rough, preliminary view on the problem and to fix ideas, 
we now give an informal presentation of the model and explain the
difficulties more concretely. The received signal is modeled as
\begin{equation}
\label{sigmod}
y(t)=\sum_{k=1}^K g_kh(t-t_k)+n(t),~~~~~~0\le t < T,
\end{equation}
where $K$ is the number of photo--electrons generated during the time interval $[0,T)$,
$h(\cdot)$ is the current pulse contributed by a single electron (which is nearly equal to the
charge of the electron multiplied by the Dirac delta 
function), $\{t_k\}$ are the random Poisson arrival times induced by
the optical signal, $\lambda(t)$, sensed by the APD, $\{g_k\}$ are the random gains induced by 
the APD photo-multiplication, and $\{n(t)\}$ is thermal noise, modeled here,
and in earlier works, to be
white Gaussian noise with spectral density $N_0/2$.\footnote{The flat spectrum
assumption is adopted here mainly for the sake of simplicity of the exposition. The extension to colored
noise is not difficult.}
The signal detection problem, in its basic form, is about
binary hypothesis testing. The null hypothesis is that $y(t)=n(t)$, whereas 
the alternative is as in (\ref{sigmod}). Had $K$, $\{g_k\}$ and $\{t_k\}$ been known to the receiver, 
the likelihood ratio (LR) would have been readily given by
(see, e.g., \cite{FGS75}):
\begin{eqnarray}
\label{lr}
L&=&\frac{\exp\left\{-\frac{1}{N_0}\int_0^T\left[y(t)-\sum_{k=1}^Kg_kh(t-t_k)\right]^2\mbox{d}t\right\}}
{\exp\left\{-\frac{1}{N_0}\int_0^Ty^2(t)\mbox{d}t\right\}}\nonumber\\
&=&\exp\left\{\frac{2}{N_0}\sum_{k=1}^K\int_0^Ty(t)h(t-t_k)\mbox{d}t-
\frac{1}{N_0}\sum_{k=1}^K\sum_{l=1}^Kg_kg_l
R(t_k-t_l)\right\},
\end{eqnarray}
where $R(\tau)=\int_0^Th(t)h(t-\tau)\mbox{d}t$.
Since these random parameters are unknown, the actual LR must be 
obtained by taking the expectation of $L$ with
respect to (w.r.t.) their randomness. Deriving this expectation appears to be notoriously difficult, 
mainly due to the second term at the exponent, i.e., the double sum over $k$ and $l$. 

It is this difficulty that
triggered many researchers in the field to harness 
their wisdom in the quest for satisfactory solutions, and
accordingly, there is rich literature on the subject,
dating back many years into the past. As far as general guidelines go, 
a possible approach to alleviate this difficulty is
the {\it estimator--correlator approach} \cite{Kailath69}, which asserts that
the expected LR of detection of a random signal in Gaussian additive white noise 
is given by the same expression as 
if the desired signal, $\sum_kg_kh(t-t_k)$, was known (i.e., the same as if
$K$, $\{g_k\}$ and
$\{t_k\}$ were known), except that it is replaced by its causal, 
minimum mean--square error (MMSE) estimator given $\{y(t)\}$. 
The caveat, however, is clear: 
deriving this MMSE causal estimator is an extremely difficult problem on its own.

To the best of the author's knowledge, the first article that is directly relevant to this kind of
study, for the above described specific signal model, 
is the article by Foschini, Gilbert and Salz \cite{FGS75}. Their approach was to view the factor
associated with the double sum in the second line of (\ref{lr}), namely, the term,
$\exp\left\{-\frac{1}{N_0}\sum_{k,l}g_kg_lR(t_k-t_l)\right\}$,
as the characteristic function of the random variable $\sum_k g_kx(t_k)$ (for given
$K$, $\{g_k\}$ and $\{t_k\}$), where $\{x(t)\}$
is an auxiliary zero--mean, stationary Gaussian process with auto-correlation function $R(\tau)$.
At the next step, the expectation over the randomness of $\{x(t)\}$ was commuted with the
expectations over $K$, $\{t_k\}$ and $\{g_k\}$, which are easier to carry out for a given realization
of $\{x(t)\}$. The result is a more compact expression of the LR, but even after this simplification, 
it is not explicit enough to be implementable in practice, or to analyze its
performance in full generality. At this point, the approach taken in \cite{FGS75} was
to carry out a series of approximations, yielding explicit 
asymptotic forms of the optimal detector and its performance at least in the 
limits of very low and very high signal--to--noise ratio (SNR). The resulting
approximate LRT for high SNR, however, was still rather complicated to implement.
Also, the behavior for moderate SNR was left open.

A year later, Mazo and Salz \cite{MS76} studied the performance of integrate--and--dump filters and
also obtained exact formulas for the random gain of the APD on the basis of the earlier study by
Personick \cite{Personick71a}, \cite{Personick71b}. See also \cite{TM94}.
Kadota \cite{Kadota88} has also derived an approximate LR test for a model like (\ref{sigmod}).
His approximation approach was different from that of \cite{FGS75}. 
It was based on neglecting the effect of overlaps between
localized noise elements, which basically amounts to ignoring the cross terms of the double summation in
the exponent of (\ref{lr}) on the ground that $R(\cdot)$ is a very narrow function (see also \cite{Hero91}
who used the same approximation for the purpose of estimation).
More recently, Helstorm and Ho \cite{HH92} and
Ho \cite{Ho95} have applied saddle--point integration and thereby 
studied the behavior of certain pulse shapes
at the optical receiver in terms of the performance of the decoder. Other studies are guided by the
approach of approximating the distribution of the shot noise of the 
photo-diode by the Gaussian distribution, owing to considerations 
in the spirit of the central limit theorem 
(CLT), see e.g., \cite[Subsections 5.6.3, 5.8.4]{Einarsson96}, \cite{EH81}. In this context,
the well--known {\it optical matched filter} 
(see, e.g., \cite{GP87}, \cite{Hero91}) is the main building block
of the optimal detector that simply maximizes the SNR at the sampling time, $t=T$.
This Gaussian approximation approach, however, raises some concerns since the CLT is not valid
for assessing the tails of the distribution and in particular, error exponents, which 
are the relevant players when probabilities of large deviations events, like (the asymptotically
rare) FA and MD error 
events, are studied.

In this paper, we take a different approach.
Motivated by considerations of the desired simplicity of
optical detectors for LIDAR 
systems (especially when they need to be implemented on mobile devices), we consider 
the class of optical signal detectors that are based on correlating the noisy received signal with
a given deterministic waveform, and we
characterize the waveform with the best trade--off between the
false--alarm (FA) probability and the missed--detection (MD) probability. 
More precisely, our derivation addresses the trade-off between 
the asymptotic error exponents of the FA and MD probabilities using
Chernoff bounds, without resorting to Gaussian approximations. We also provide numerical 
results that compare the performance of the best correlator to that of
the optical matched filter (or, more precisely, the matched correlator), 
which is coherent with the above--mentioned Gaussian approximation approach.
It is demonstrated that the proposed optimal 
correlator outperforms the optical matched correlator, in terms of the
trade--off between the FA and the MD error exponents.
It should be pointed out that in addition to the random fluctuations of the
APD photo-multiplier, our model also incorporates the effect of 
dark current that exists even under the null hypothesis.

In the same spirit and with a similar motivation, 
we also study, for the same type of signal model, the
problem of estimating the delay of a received signal 
on the basis of maximizing the correlation between the received signal and
a time--shifted waveform, as a function 
of this shift. We characterize the optimal correlator waveform
that minimizes the mean square error (MSE) 
in the regime of high SNR, as an extension of the analysis
provided by Bar-David \cite{BarDavid69}, who analyzed the high--SNR MSE of the
maximum likelihood (ML) estimator
for the pure Poissonian regime (i.e., without thermal noise).
Once again, the emphasis is on simplicity and 
therefore, the performance of this estimator cannot be compared
to the much more complicated, approximate MAP estimator due to Hero \cite{Hero91}, which is based on 
approximating the likelihood function, using the same approach as Kadota \cite{Kadota88}.

The optimal correlator waveforms for detection and for estimation turn out to be different,
but their limiting behavior is the same in 
both detection and estimation problems:
when the thermal Gaussian noise is dominant,
the optimal correlator waveform becomes proportional to 
the clean signal (like the classical matched filter for
additive white Gaussian noise), but when the thermal noise is
negligible compared to the other noises, then it becomes logarithmic 
function of the clean signal, as expected in view of \cite{BarDavid69}.

The outline of the remaining part of the paper is as follows.
In Section \ref{thesignalmodel}, we present the model under discussion in full detail.
In Section \ref{detection}, we address the signal 
detection problem, first and foremost, for the case of zero dark current. The case of positive
dark current, which follows the same general ideas
(but more complicated), is also outlined, but relatively briefly.
Finally, in Section \ref{estimation}, we address the problem of time delay estimation.

\section{The Signal Model}
\label{thesignalmodel}

In this section, we provide a formal presentation of the signal model, 
that was briefly described in the Introduction. As mentioned before,
we consider the model,
\begin{equation}
\label{model}
y(t)=\sum_{k=1}^K g_k h(t-t_k)+n(t),~~~~~~~~~~~~0\le t < T,
\end{equation}
whose various ingredients are described as follows. The variable
$K$ is a Poissonian random variable, distributed according to
\begin{equation}
\mbox{Pr}\{K=\kappa\}=e^{-\Lambda}\frac{\Lambda^\kappa}{\kappa!},
~~~~~~\kappa=0,1,2\ldots,~~~~\Lambda=\int_0^T\lambda(t)\mbox{d}t,
\end{equation}
where $\lambda(t)$ is the a rate function 
that depends upon the intensity of the received optical signal. In particular, 
\begin{equation}
\lambda(t)=\frac{\eta\calP(t)}{\hbar\omega}+\lambda_{\mbox{\tiny d}}, 
\end{equation}
where $\eta$ is the quantum efficiency of the APD, $\calP(t)$ is the
instantaneous power of the optical signal, 
$\hbar$ is Planck's constant, $\omega$ is the angular frequency of the light wave, and
$\lambda_{\mbox{\tiny d}}$ is the dark current. The variables
$\{g_k\}$ are independently identically distributed (i.i.d.) positive integer 
random variables that designate the avalanche gains. According to Personick \cite{Personick71a}, 
\cite{Personick71b}, the distribution of these random variables depends on
the physics of the APD, and its characteristic function obeys a certain implicit equation,
which is solvable in closed form when only the 
electrons (and not holes) cause ionizing collisions. In this case,
the distribution of each $g_k$ is geometric:
\begin{equation}
\label{geom}
\mbox{Pr}\{g_k=g\}=(e^\zeta-1)\cdot e^{-\zeta g},~~~~~~g=1,2,\ldots,~~\zeta > 0.
\end{equation}
For the sake of concreteness, we will henceforth adopt the assumption of this geometric distribution.
It also includes the case of a deterministic gain ($g_k =1$ with probability
one), which corresponds to the case of the PIN diode, by taking the limit
$\zeta\to\infty$.
The function $h(t)$ is the current pulse contributed
by the passage of a single photo--electron and hence its integral must be equal to the
electric charge of the electron, $q_{\mbox{\tiny e}}$. Naturally, 
this is a very narrow pulse, which for most practical purposes, can be approximated by $h(t)\approx
q_{\mbox{\tiny e}}\delta(t)$, where $\delta(t)$ is the Dirac delta function. However, $h(t)$ can also be
understood to include the convolution with some front--end 
filter, which is part of the electronic circuitry (e.g., the TIA). 
The times $\{t_k\}$ are the random Poissonian
photon arrival times, taking on values in $[0,T)$ and being induced by the optical waveform, $\lambda(t)$.
Finally, $\{n(t)\}$ is Gaussian white noise
with spectral density $N_0/2$, which is assumed to be independent of $K$, $\{g_k\}_{k=1}^K$ and
$\{t_k\}_{k=1}^K$. Also, given $K$, $\{g_k\}$ are statistically independent of $\{t_k\}$. Recall that for
the underlying Poissonian process defined, conditioned on the event $K=\kappa$, the 
unordered random arrival times,
$t_1,\ldots,t_\kappa$, are i.i.d.\ and their common density function 
is given by $f(t)=\lambda(t)/\Lambda$, for $0\le t\le T$, and $f(t)=0$ elsewhere.

Observe that if we present $g_k$ as $\bar{g}+\Delta g_k$, where
$\bar{g}=\bE\{g_k\}$ and then $\Delta g_k$ designates the fluctuation, 
then the random input signal can be represented as
\begin{eqnarray}
\sum_k(\bar{g}+\Delta g_k)h(t-t_k)&=&\bar{g}\sum_kh(t-t_k)+\sum_k\Delta
g_kh(t-t_k)\nonumber\\
&=&\bar{g}[q_{\mbox{\tiny e}}\lambda(t)+n_{\mbox{\tiny s}}(t)]+\sum_k\Delta
g_kh(t-t_k)\nonumber\\
&=&\bar{g}q_{\mbox{\tiny e}}\lambda(t)+\bar{g}n_{\mbox{\tiny
s}}(t)+\sum_k\Delta g_kh(t-t_k).
\end{eqnarray}
The first term, $\bar{g}q_{\mbox{\tiny e}}\lambda(t)$, is the desired clean
signal (plus dark current), the second term, defined as
\begin{equation}
\bar{g}n_{\mbox{\tiny s}}(t)=\bar{g}\left[\sum_kh(t-t_k)-q_{\mbox{\tiny e}}\lambda(t)\right],
\end{equation}
is shot noise (amplified by $\bar{g}$), and the last
term is multiplicative noise. Thus, together with the Gaussian noise, $n(t)$,
of (\ref{model}), there are three types of noise in this model, as already mentioned in the Introduction.

\section{Signal Detection}
\label{detection}

In this section, we study the signal detection problem.
We begin with the case $\lambda_{\mbox{\tiny d}}=0$ (no dark current), 
which is considerably simpler, and then
outline the extension to the more general case,
$\lambda_{\mbox{\tiny d}} > 0$. Before, we move into the technical details, a comment is in order,
and it applies even to the case of no dark current.
Consider then the signal detection problem of deciding between the two
hypotheses:
\begin{eqnarray}
& &\calH_0:~y(t)=n(t)\\
& &\calH_1:~y(t)=\sum_{k=1}^K g_k h(t-t_k)+n(t),
\end{eqnarray}
using a detector, that is based on a correlator, that is, calculating the quantity
$$\int_0^T w(t)y(t)\mbox{d}t$$
and comparing it to a threshold, $\theta T$. Here,
$\{w(t),~0\le t\le T\}$ is a deterministic waveform to be optimized,
and $\theta > 0$ is a threshold parameter that controls the trade-off between the
FA and MD probabilities. 
Clearly, if 
the noise was purely Gaussian white noise, 
under both hypotheses, the optimal choice of $w(t)$ would
have been matched to the desired signal, 
i.e., $w(t)=\lambda(t)$. Here, however, as explained in Section
\ref{thesignalmodel}, there are two additional types of non--Gaussian noise under $\calH_1$.
Since the classical matched correlator, $w(t)=\lambda(t)$, is no longer 
necessarily optimal under non--Gaussian noise, and since we would
still be interested in a detector that is relatively 
easy to implement, the natural question is whether
there is a waveform $w(t)$ better than $w(t)=\lambda(t)$. If so, then what is the optimal waveform $w_*(t)$
for this detection problem? 
In this context, we should also mention again the notion of the optical matched filter (see
e.g., \cite{GP87}, \cite{Hero91}), the optical analogue of the classical matched filter,
which maximizes the SNR at the sampling time, $t=T$, taking
into account that the intensity of the shot noise is proportional to the
desired signal, $\lambda(t)$ (unlike the case of pure thermal noise). 
But the relevance of the SNR as the only
parameter that counts for detection performance is valid 
only under the Gaussian regime, so the optical matched filter
is applied either under the Gaussian approximation, or when only second moments are important. 
The Gaussian approximation, under this model, 
is largely justified by CLT considerations. Here, however, we wish
to avoid CLT considerations, as we are interested in the FA and MD error exponents, which are,
in fact, given by large deviations rate functions.
As is well known, the CLT is not valid for the large-deviations regime and for tails 
of distributions.
Indeed, the optimal $w_*(t)$ that we derive below will be different from the optical matched filter.

\subsection{The Case of No Dark Current}

Consider the hypothesis testing problem defined at the beginning of the 
introductory part of this section, which
is for the case of no dark current.
Let $E\dfn\int_0^Tw^2(t)\mbox{d}t$ and $P\dfn E/T$. Then,
the FA probability is given by
\begin{eqnarray}
P_{\mbox{\tiny FA}}&=&\mbox{Pr}\left\{\int_0^T w(t)n(t)\mbox{d}t > \theta
T\right\}\nonumber\\
&=&Q\left(\frac{\theta T}{\sqrt{N_0E/2}}\right)\nonumber\\
&\exe&\exp\left\{-\frac{\theta^2T^2}{N_0E}\right\}\nonumber\\
&=&\exp\left\{-\frac{\theta^2T}{N_0P}\right\}\nonumber\\
&\dfn&e^{-E_{\mbox{\tiny FA}}(\theta)T}
\end{eqnarray}
where the notation $\exe$ designates asymptotic equivalence 
in the exponential scale, i.e., for two positive functions of $T$, $a(T)$ and $b(T)$,
the assertion $a(T)\exe b(T)$, 
stands for the assertion that
$\lim_{T\to\infty}\frac{1}{T}\log\frac{a(T)}{b(T)}=0$.
Since the FA error exponent, $E_{\mbox{\tiny FA}}(\theta)=\theta^2/N_0P$, 
depends on the waveform $w(\cdot)$ only via its
power, $P$, it is obvious that the maximization of the MD error exponent for given
FA error exponent is equivalent to its maximization subject to a power
constraint imposed on $w(\cdot)$.
Denoting $Z\dfn\int_0^T w(t)n(t)\mbox{d}t$, we next assess the MD error exponent using the Chernoff bound,
assuming that $h(t)=q_{\mbox{\tiny e}}\delta(t)$. 
\begin{eqnarray}
P_{\mbox{\tiny MD}}&=&\mbox{Pr}\left\{\int_0^T w(t)y(t)\mbox{d}t \le \theta
T\right\}\nonumber\\
&=&\mbox{Pr}\left\{q_{\mbox{\tiny e}}\sum_{k=1}^Kg_kw(t_k)+Z \le \theta
T\right\}\nonumber\\
&\le&\inf_{s\ge 0} e^{s\theta
T}\bE\left\{\exp\left[-sq_{\mbox{\tiny e}}\sum_{k=1}^Kg_kw(t_k)-sZ\right]\right\}\nonumber\\
&=&\inf_{s\ge 0} \exp\left\{s\theta
T+\frac{s^2N_0PT}{4}\right\}\bE\left\{\exp
\left[-sq_{\mbox{\tiny e}}\sum_{k=1}^Kg_kw(t_k)\right]\right\}\nonumber\\
&=&\inf_{s\ge 0} \exp\left\{s\theta
T+\frac{s^2N_0PT}{4}\right\}\bE\left\{\prod_{k=1}^Ke^{-sq_{\mbox{\tiny e}}g_kw(t_k)}\right\}.
\end{eqnarray}
To calculate the last expectation, we proceed in two steps. First, we average
each factor over $g_k$, assuming that it is geometrically distributed as in (\ref{geom}).
This gives
\begin{eqnarray}
\sum_{g=1}^\infty P(g)e^{-gsq_{\mbox{\tiny e}}w(t_k)}&=&(e^\zeta-1)
\sum_{g=1}^\infty e^{-g[sq_{\mbox{\tiny e}}w(t_k)+\zeta]}\nonumber\\
&=&\frac{e^{\zeta}-1}{e^{sq_{\mbox{\tiny e}}w(t_k)+\zeta}-1}.
\end{eqnarray}
As a second step, we average over $K$ and $\{t_k\}_{k=1}^K$, and get
\begin{eqnarray}
\bE\left\{\prod_{k=1}^Ke^{-sq_{\mbox{\tiny e}}g_kw(t_k)}\right\}&=&\bE\left\{\prod_{k=1}^K\frac{e^{\zeta}-1}
{e^{sq_{\mbox{\tiny e}}w(t_k)+\zeta}-1}\right\}\nonumber\\
&=&\exp\left\{\int_0^T\lambda(t)\left[\frac{e^{\zeta}-1}
{e^{sq_{\mbox{\tiny e}}w(t)+\zeta}-1}-1\right]\mbox{d}t\right\}\nonumber\\
&=&\exp\left\{-e^{\zeta}\int_0^T\lambda(t)\cdot\frac{e^{sq_{\mbox{\tiny e}}w(t)}-1}
{e^{sq_{\mbox{\tiny e}}w(t)+\zeta}-1}\mbox{d}t\right\},
\end{eqnarray}
where we have used the fact 
\cite[eqs.\ (8)--(13)]{BarDavid69} that for an arbitrary positive function $f$,
\begin{equation}
\bE\left\{\prod_{k=1}^K f(t_k)\right\}=\exp\left\{\int_0^T\lambda(t)[f(t)-1]\mbox{d}t\right\}.
\end{equation}
Thus,
\begin{equation}
P_{\mbox{\tiny MD}}\le
\exp\left\{-T\sup_{s\ge
0}\left[\frac{e^{\zeta}}{T}\int_0^T\lambda(t)\cdot\frac{e^{sq_{\mbox{\tiny e}}w(t)}-1}
{e^{sq_{\mbox{\tiny e}}w(t)+\zeta}-1}\mbox{d}t-s\theta-s^2\frac{N_0P}{4}\right]\right\}.
\end{equation}
Consider first the special case where $g_k\equiv 1$ with probability one,
which is obtained in the limit
$\zeta\to\infty$. In this case, the above simplifies to
\begin{equation}
P_{\mbox{\tiny MD}}\le
\exp\left\{-T\sup_{s\ge 0}\left[\frac{1}{T}\int_0^T\lambda(t)[1-e^{-sq_{\mbox{\tiny e}}w(t)}]
\mbox{d}t-s\theta-s^2\frac{N_0P}{4}\right]\right\}.
\end{equation}

\noindent
{\it Remark.} Note that had the channel been purely Gaussian, that is, without the shot noise, and the
desired signal was $q_{\mbox{\tiny e}}\lambda(t)$, we would have obtained
\begin{equation}
P_{\mbox{\tiny MD}}\le
\exp\left\{-T\sup_{s\ge 0}\left[\frac{1}{T}\int_0^Tq_{\mbox{\tiny e}}\lambda(t)\cdot sw(t)
\mbox{d}t-s\theta-s^2\frac{N_0P}{4}\right]\right\}.
\end{equation}
This means that the difference
$$\frac{1}{T}\int_0^T\lambda(t)\cdot[sq_{\mbox{\tiny e}}w(t)-(1-e^{-sq_{\mbox{\tiny e}}w(t)})]\mbox{d}t$$
designates the loss due to the additional shot noise.\footnote{The integrand
is, of course, non--negative due to the inequality $x\ge 1-e^{-x}$.}

We therefore need to maximize the exponent over both $s$ and $\{w(t),~0\le t\le T\}$.
For a given $s$, the optimal $\{w(t)\}$ minimizes
$\int_0^T\lambda(t)e^{-sq_{\mbox{\tiny e}}
w(t)}\mbox{d}t$ subject to the power constraint
$\int_0^Tw^2(t)\mbox{d}t\le PT$, which is equivalent to minimizing
\begin{equation}
\int_0^T\lambda(t)e^{-sq_{\mbox{\tiny e}}w(t)}\mbox{d}t+
\frac{s^2q_{\mbox{\tiny e}}^2}{2c}\left[\int_0^Tw^2(t)\mbox{d}t-PT\right],
\end{equation}
where $s^2q_{\mbox{\tiny e}}^2/2c > 0$ is a Lagrange 
multiplier.\footnote{This form of the Lagrange multiplier is adopted as it allows a convenient
representation of the solution. It is legitimate since there is complete freedom to control
its value by the choice of the constant $c > 0$.}
Finding the optimal function $w(\cdot)$ is a standard problem in
calculus of variations, whose solution is characterized as follows.
Let the function $p[\cdot]$ denote the inverse of the monotonically increasing function
$b[x]\dfn xe^x$, $x\ge 0$, i.e., $p[y]$ is the solution $x$ to the equation $xe^x=y$,
$y\ge 0$. The optimal $w(t)$ is given by
\begin{equation}
w_*(t)=\frac{1}{sq_{\mbox{\tiny e}}}p[c\cdot\lambda(t)],
\end{equation}
where $c > 0$ is chosen such that
$\int_0^Tp^2[c\cdot\lambda(t)]\mbox{d}t=s^2q_{\mbox{\tiny e}}^2PT$.
The MD exponent is then given by
\begin{equation}
E_{\mbox{\tiny MD}}(\theta)=\sup_{s\ge
0}\left[\frac{1}{T}\int_0^T\lambda(t)(1-\exp\{-p[c\cdot\lambda(t)]\})
\mbox{d}t-s\theta-s^2\frac{N_0P}{4}\right],
\end{equation}
where it should be kept in mind that $c$ depends on $s$.

Note that if the optimal $s$ is very small (which is the case when $N_0$ and/or $\theta$
are large, then the solution to the equation 
$sq_{\mbox{\tiny e}}w(t)e^{sq_{\mbox{\tiny e}}w(t)}=c\cdot\lambda(t)$
is found near the origin, where $p[x]\approx x$, which means that $w_*(t)$ is
nearly proportional to $\lambda(t)$, namely, the classical matched correlator. If, on
the other hand, the optimal $s$ is very large (which is the case when $N_0$ and $\theta$ are both
small), then the solution is found away
from the origin, where $p[x]\approx \ln x$. In this case,
$w_*(t)\approx\frac{1}{sq_{\mbox{\tiny e}}}\ln[c\cdot\lambda(t)]$, in agreement of optimal
photo--counting detector (see, e.g., \cite{BarDavid69}), 
which is obtained in the absence of Gaussian noise.\\

\begin{figure}[h!t!b!]
\centering
\includegraphics[width=8.5cm, height=8.5cm]{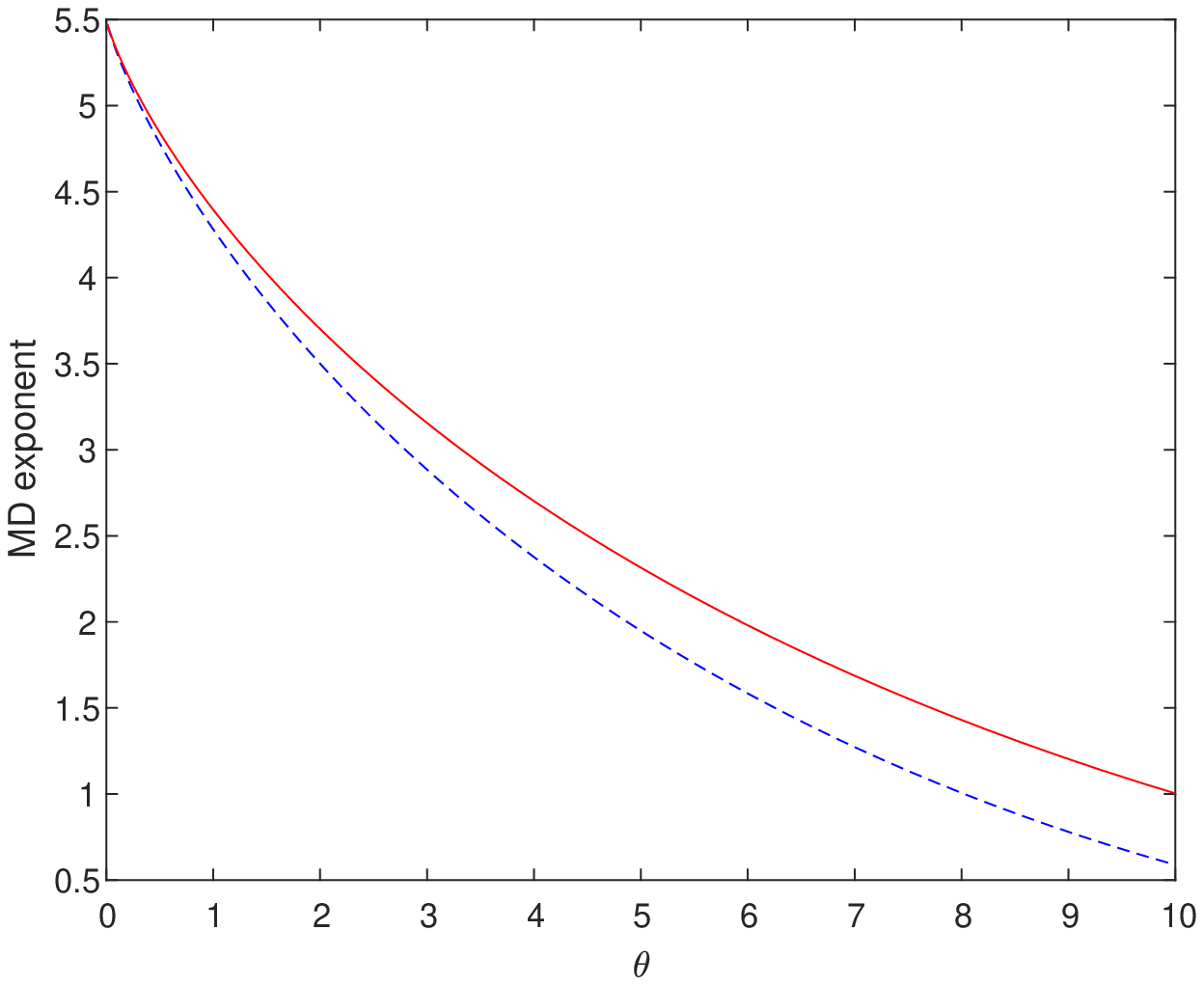}
\caption{MD error exponents of the optical matched correlator (dashed blue
curve) and the
optimal correlator $w_*$ (solid red curve) as functions of 
$\theta$, for a deterministic gain
($\zeta\to\infty$), $P=10$,
$N_0/q_{\mbox{\tiny e}}^2=0.0001$, $\lambda_1=1$ and $\lambda_2=10$.}
\label{graph1}
\end{figure}

\noindent
{\it Example 1.}
Consider the frequently--encountered case where $\lambda(t)$ is a two--level signal,
where half of the time $\lambda(t)=\lambda_1$ and in the other half,
$\lambda(t)=\lambda_2$. Then, $w_*(t)$ must also be a two--level signal. Owing to the power constraint, 
we may denote these two levels by $w$ and $\sqrt{2P-w^2}$, respectively, and it remains to maximize the
exponent over $s$ and $w$ alone. Specifically, we have
\begin{equation}
E_{\mbox{\tiny MD}}(\theta)=\sup_{s\ge
0}\max_{0\le w\le\sqrt{2P}}\left[\frac{\lambda_1}{2}(1-e^{-sq_{\mbox{\tiny e}}w})+
\frac{\lambda_2}{2}(1-e^{-sq_{\mbox{\tiny e}}\sqrt{2P-w^2}})
-s\theta-s^2\frac{N_0P}{4}\right].
\end{equation}
Fig.\ \ref{graph1} compares the exponent of the optimal waveform, $w_*$, to that of
the optical matched filter \cite[eq.\ (8)]{GP87},
\begin{equation}
w_{\mbox{\tiny omf}}(t)=\frac{\lambda(t)}{\lambda(t)+N_0/(2q_{\mbox{\tiny e}}^2\overline{g^2})},
\end{equation}
for the following values of the parameters of the problem: 
$P=10$, $N_0/q_{\mbox{\tiny e}}^2=0.0001$, $\lambda_1=1$ and $\lambda_2=10$.
The numerical value of the spectral density 
of the thermal noise was deliberately chosen extremely small, in order
to demonstrate a situation where the noise is far from being Gaussian, and
thereby examine sharply the validity of the Gaussian approximation. In
this case, $w_{\mbox{\tiny omf}}(t)$ is nearly equal to unity for all $t$,
which means a pure, unweighted integrator \cite[p.\ 1291, Remark 2]{GP87}.
As can be seen, the optimal correlator $w_*$ improves upon the optical matched
correlator fairly significantly, especially for large values of the threshold
parameter, $\theta$. This concludes Example 1. $\Box$

Returning to the case of a general, finite $\zeta$, 
and carrying out a similar optimization, we find that
the optimal $w_*(t)$ is now given by
\begin{equation}
w_*(t)=\frac{1}{sq_{\mbox{\tiny e}}}\cdot p_\zeta[c\cdot\lambda(t)], 
\end{equation}
where $p_\zeta$ is the inverse of the function
\begin{equation}
b_\zeta[x]=\frac{x(e^{x+\zeta}-1)^2}{e^{x+2\zeta}-e^{x+\zeta}},
\end{equation}
and where, once again, $c$ is chosen such that
\begin{equation}
\int_0^Tp_\zeta^2[c\cdot\lambda(t)]\mbox{d}t=s^2q_{\mbox{\tiny e}}^2PT.
\end{equation}
Here too, if $N_0$ and/or $\theta$ are large, then $s$ must be small, and then
due to the power constraint, $c$ must be small too, which means that the
functions $p_\zeta$ and $b_\zeta$ operate near the origin, 
where they are roughly linear, as
\begin{equation}
b_\zeta[x]\approx x\cdot\frac{(e^\zeta-1)^2}{e^{2\zeta}-e^\zeta}.
\end{equation}
At the other extreme, on the other hand, $p_\zeta$ and $b_\zeta$ operate away from the origin,
where
\begin{equation}
b_\zeta[x]\approx
\frac{xe^{2x+2\zeta}}{e^{x+2\zeta}-e^{x+\zeta}}=\frac{xe^{x+\zeta}}{e^\zeta-1},
\end{equation}
which is again, nearly exponential, and so, $p_\zeta$ is approximately logarithmic,
as before. The MD exponent is therefore given by
\begin{equation}
E_{\mbox{\tiny MD}}(\theta)=
\sup_{s\ge
0}\left[\frac{e^{\zeta}}{T}\int_0^T\lambda(t)\cdot\frac{\exp\{p_\zeta[c\cdot\lambda(t)]\}-1}
{\exp\{p_\zeta[c\cdot\lambda(t)]+\zeta\}-1}\mbox{d}t-s\theta-s^2\frac{N_0P}{4}\right].
\end{equation}

\noindent
{\it Example 2.} Consider again the setting of Example 1 above, except that here $\zeta$ is finite.
Specifically, Fig.\ 2 displays a comparison analogous to that of Fig.\ 1
for the case of a random gain with
parameter $\zeta=0.1$, where for the red solid curve, $s$ and $w$ are chosen to maximize
\begin{equation}
E_{\mbox{\tiny MD}}(\theta)=\sup_{s\ge
0}\max_{0\le
w\le\sqrt{2P}}\left[\frac{\lambda_1e^\zeta}{2}\cdot\frac{e^{sq_{\mbox{\tiny e}}w}-1}
{e^{sq_{\mbox{\tiny e}}w+\zeta}-1}+
\frac{\lambda_2e^\zeta}{2}\cdot\frac{e^{sq_{\mbox{\tiny e}}\sqrt{2P-w^2}}-1}
{e^{sq_{\mbox{\tiny e}}\sqrt{2P-w^2}+\zeta}-1}
-s\theta-s^2\frac{N_0P}{4}\right],
\end{equation}
and for the blue, dashed 
curve, $w$ is chosen to be the optical matched correlator as before.
As can be seen, here too, the optimal $w$ improves upon the optical matched
correlator and the gap is rather considerable, especially as $\theta$ grows.

\begin{figure}[h!t!b!]
\centering
\includegraphics[width=8.5cm, height=8.5cm]{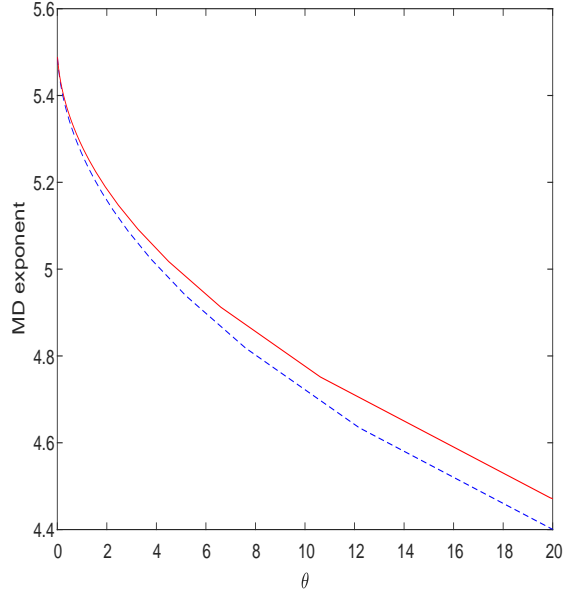}
\caption{MD error exponents of the optical matched correlator (blue dashed
curve) and
the optimal correlator (red solid curve) as functions of $\theta$, for a random gain
($\zeta=0.1$), $P=10$,
$N_0/q_{\mbox{\tiny e}}^2=0.0001$, $\lambda_1=1$ and $\lambda_2=10$.}
\label{graph2}
\end{figure}

\subsection{The Case of Positive Dark Current}

The case of $\lambda_{\mbox{\tiny d}} > 0$ is analyzed on the basis of similar ideas, and we
therefore cover it relatively briefly, highlighting mostly the points where there is a substantial
difference relative to the zero dark--current case.

Extending the analysis to the case of positive dark current, a similar
derivation yields the following FA exponent for a given correlator waveform, $w(t)$:
\begin{equation}
E_{\mbox{\tiny FA}}(\theta,w)=\sup_{0\le s<
\zeta/w_{\max}}\left[s\theta-\frac{s^2N_0P}{4}-\frac{\lambda_{\mbox{\tiny d}}
e^\zeta}{T}\int_0^T\frac{e^{sq_{\mbox{\tiny e}}w(t)}-1}{e^\zeta-e^{sq_{\mbox{\tiny e}}w(t)}}\mbox{d}t\right],
\end{equation}
where $w_{\max}=\sup_{0\le t\le T}w(t)$.
Here, the trade--off between the FA and the MD error exponents is somewhat more
involved than in the zero dark--current case, 
since the FA error exponent depends on $w(\cdot)$ in a more complicated manner
than just via its power $P$.
In particular, we would now like
to find the pair $(\theta,w)$ that maximizes $E_{\mbox{\tiny MD}}(\theta,w)$ over all pairs
such that $E_{\mbox{\tiny FA}}(\theta,w)\ge S$, for a given $S >
0$ that designates the target FA exponent.\footnote{Previously, 
$S$ was given by $\theta^2/N_0P$, so for a given $P$, $S$ was
proportional to $\theta^2$.} Equivalently, we would like to solve the problem
\begin{equation}
\sup_{\theta,w}\left\{E_{\mbox{\tiny
MD}}(\theta,w)+\mu E_{\mbox{\tiny FA}}(\theta,w)\right\},
\end{equation}
where $\mu$ is a Lagrange multiplier chosen to meet the constraint, $E_{\mbox{\tiny FA}}(\theta,w)\ge S$.
More specifically, using Chernoff bounds for both error exponents, this
amounts to solving the problem,
\begin{eqnarray}
& &\sup_{s\ge 0}\sup_{\sigma\ge 0}\sup_{\theta\ge 0}\sup_{w:~w_{\max} < \zeta/s}\left[
\frac{e^{\zeta}}{T}\int_0^T\lambda(t)\cdot\frac{e^{\sigma  q_{\mbox{\tiny e}} w(t)}-1}
{e^{\sigma q_{\mbox{\tiny e}}
w(t)+\zeta}-1}\mbox{d}t-\sigma\theta-\sigma^2\frac{N_0P}{4}+\right.\nonumber\\
& &\left.\mu s\theta-\frac{\mu s^2N_0P}{4}-\frac{\mu\lambda_0
e^\zeta}{T}\int_0^T\frac{e^{sq_{\mbox{\tiny e}}w(t)}-1}{e^\zeta-
e^{sq_{\mbox{\tiny e}}w(t)}}\mbox{d}t\right]\nonumber\\
&=&\sup_{s\ge 0}\sup_{\sigma\ge 0}\sup_{\theta\ge 0}\sup_{w:~w_{\max} <
\zeta/s}\left[
\frac{e^{\zeta}}{T}\int_0^T\left(\lambda(t)\cdot\frac{e^{\sigma q_{\mbox{\tiny e}}w(t)}-1}
{e^{\sigma q_{\mbox{\tiny e}}
w(t)+\zeta}-1}-\mu\lambda_0
\frac{e^{sq_{\mbox{\tiny e}}w(t)}-1}{e^\zeta-e^{sq_{\mbox{\tiny e}}w(t)}}\right)\mbox{d}t+\right.\nonumber\\
& &\left.(\mu s-\sigma)\theta-(\sigma^2+\mu s^2)\frac{N_0P}{4}\right].
\end{eqnarray}
We will not continue any further to the full, detailed solution of this
problem, beyond the
following comment which applies to the case of a deterministic gain.
In the limit of $\zeta\to\infty$, the above trade-off yields
\begin{equation}
w_*(t)=\ln\left[\frac{\sigma}{\lambda_{\mbox{\tiny d}}\mu s}\cdot\lambda(t)\right],
\end{equation}
in other words, in the presence of dark-current, the relation is always
logarithmic.

\section{Time Delay Estimation}
\label{estimation}

In this section, we consider the problem of time 
delay estimation. The underlying model is the same as
before, except that the optical signal 
is time--shifted, i.e., $\lambda(t-\theta)$, where $\theta$ is the
delay. It is assumed that the support of 
$\lambda(t-\theta)$ in included in the interval $[0,T]$, for every
$\theta$ in the range of uncertainty.
As mentioned in the Introduction, this is a relevant problem in LIDAR systems, where distances to
certain objects have to be estimated, similarly as in classical radar systems.
Here too, our basic building block is a correlator. Consider an estimator of the
form,
\begin{equation}
\hat{\theta}=\arg\max_\theta Q(\theta),
\end{equation}
where
\begin{equation}
Q(\theta)=\int_0^Ty(t)w(t-\theta)\mbox{d}t,
\end{equation}
and where $w(\cdot)$ is a twice--differentiable waveform to be optimized, with the property
that the temporal cross--correlation function,
\begin{equation}
R_{w\lambda}(\tau)=\int_0^T\lambda(t)w(t-\tau)\mbox{d}t
\end{equation}
achieves its maximum at $\tau=0$. This implies
\begin{equation}
\dot{R}_{w\lambda}(0)=\int_0^T\lambda(t)\dot{w}(t)\mbox{d}t=0,
\end{equation}
where dotted functions designate derivatives. Similarly as the assumption concerning $\lambda(\cdot)$, 
we also assume that the support of the waveform
$w(t-\theta)$ is fully included in $[0,T]$ for the entire range of search of the estimated delay.

Our analysis begins similarly as in \cite{BarDavid69},
which assumes high SNR and small estimation errors. Accordingly, consider the
Taylor series expansion of first derivative, $\dot{Q}(\theta)$, around the
true parameter value, $\theta_0$:
\begin{equation}
0=\dot{Q}(\hat{\theta})\approx\dot{Q}(\theta_0)+(\hat{\theta}-\theta_0)\ddot{Q}(\theta_0),
\end{equation}
where $\ddot{Q}(\theta)$ is the second derivative of $Q(\theta)$. This yields
\begin{eqnarray}
\label{approxerror}
\hat{\theta}-\theta_0&\approx&-\frac{\dot{Q}(\theta_0)}{\ddot{Q}(\theta_0)}\nonumber\\
&=&-\frac{\bar{g}q_{\mbox{\tiny
e}}\int_0^T\lambda(t-\theta_0)\dot{w}(t-\theta_0)\mbox{d}t+\int_0^Tn_T(t)\dot{w}(t-\theta_0)\mbox{d}t}
{\bar{g}q_{\mbox{\tiny e}}\int_0^T\lambda(t-\theta_0)\ddot{w}(t-\theta_0)\mbox{d}t
+\int_0^Tn_T(t)\ddot{w}(t-\theta_0)\mbox{d}t}\nonumber\\
&=&-\frac{\bar{g}q_{\mbox{\tiny
e}}\int_0^T\lambda(t)\dot{w}(t)\mbox{d}t+\int_0^Tn_T(t)\dot{w}(t-\theta_0)\mbox{d}t}
{\bar{g}q_{\mbox{\tiny e}}\int_0^T\lambda(t)\ddot{w}(t)\mbox{d}t
+\int_0^Tn_T(t)\ddot{w}(t-\theta_0)\mbox{d}t}\nonumber\\
&=&-\frac{
\int_0^Tn_T(t)\dot{w}(t-\theta_0)\mbox{d}t}
{\bar{g}q_{\mbox{\tiny e}}\cdot\int_0^T\lambda(t)\ddot{w}(t)\mbox{d}t
+\int_0^Tn_T(t)\ddot{w}(t-\theta_0)\mbox{d}t}\nonumber\\
&\dfn&\frac{U_n}{A+U_d},
\end{eqnarray}
where $n_T(t)$ is the total noise,
composed of the shot noise, the multiplicative noise and the thermal noise,
i.e.,
\begin{equation}
n_{\mbox{\tiny T}}(t)=n_{\mbox{\tiny s}}(t)+n_{\mbox{\tiny m}}(t)+n(t),
\end{equation}
where
\begin{equation}
n_{\mbox{\tiny s}}(t)=\bar{g}\cdot\left[\sum_kh(t-t_k)-q_{\mbox{\tiny
e}}\lambda(t-\theta)\right],
\end{equation}
and
\begin{equation}
n_{\mbox{\tiny m}}(t)=\sum_k\Delta g_kh(t-t_k).
\end{equation}
Thus,
\begin{equation}
\bE\{(\hat{\theta}-\theta)^2\}\approx\frac{\bE\{U_n^2\}}{A^2},
\end{equation}
where we have made a further approximation by neglecting the contribution of
the random variable $U_d$ relative to the deterministic constant $A$
(at the denominator of the last line of (\ref{approxerror})) since $A$ is proportional to $T$, whereas
the standard deviation of $U_d$ is proportional 
to $\sqrt{T}$ (see also \cite[eqs.\ (38)--(40)]{BarDavid69} for a more
detailed justification of a similar approximation).
It is easy to see that all three noise components are uncorrelated with each other.
The auto-correlation function of the thermal noise is
$R_n(t,s)=\frac{N_0}{2}\delta(t-s)$. The auto-correlation function of the (amplified) shot noise is
\begin{eqnarray}
R_{\mbox{\tiny
s}}(t,s)&=&\bar{g}^2\cdot\bE\left\{\left[\sum_kh(t-t_k)-q_{\mbox{\tiny
e}}\lambda(t)\right]\cdot\left[\sum_kh(s-t_k)-q_{\mbox{\tiny
e}}\lambda(s)\right]\right\}\nonumber\\
&=&\bar{g}^2\cdot\bE\left\{\sum_{k,l}h(t-t_k)h(s-t_l)\right\}-\bar{g}^2q_{\mbox{\tiny
e}}^2\lambda(t)\lambda(s)\nonumber\\
&=&\bar{g}^2\cdot\bE\left\{\sum_{k}h(t-t_k)h(s-t_k)\right\}+
\bar{g}^2\cdot\bE\left\{\sum_{k\ne l}h(t-t_k)h(s-t_l)\right\}
-\bar{g}^2q_{\mbox{\tiny
e}}^2\lambda(t)\lambda(s)\nonumber\\
&=&\bar{g}^2\cdot\bE\{K\}\frac{1}{\Lambda}
\int_0^T\lambda(\tau-\theta)h(t-\tau)h(s-\tau)\mbox{d}\tau
+\nonumber\\
& &\bar{g}^2\cdot\bE\{K(K-1)\}\frac{1}{\Lambda^2}
\left[\int_0^T\lambda(\tau-\theta)h(t-\tau)\mbox{d}\tau\right]
\left[\int_0^T\lambda(\tau-\theta)h(s-\tau)\mbox{d}\tau\right]-\bar{g}^2q_{\mbox{\tiny
e}}^2\lambda(t)\lambda(s)\nonumber\\
&=&\bar{g}^2\cdot\int_0^T\lambda(\tau-\theta)h(t-\tau)h(s-\tau)\mbox{d}\tau.
\end{eqnarray}
Similarly,
\begin{eqnarray}
R_{\mbox{\tiny
m}}(t,s)&=&\bE\{n_{\mbox{\tiny m}}(t)n_{\mbox{\tiny m}}(s)\}\nonumber\\
&=&\bE\left\{\sum_{k,l}\Delta g_k\Delta g_lh(t-t_k)h(s-t_l)\right\}\nonumber\\
&=&\mbox{Var}\{g\}\cdot\bE\left\{\sum_{k}h(t-t_k)h(s-t_k)\right\}+
\bE\left\{\sum_{k\ne l}\Delta g_k\Delta g_lh(t-t_k)h(s-t_l)\right\}\nonumber\\
&=&\mbox{Var}\{g\}\cdot\bE\left\{\sum_{k}h(t-t_k)h(s-t_k)\right\}\nonumber\\
&=&\mbox{Var}\{g\}\cdot\int_0^T\lambda(\tau-\theta)h(t-\tau)h(s-\tau)\mbox{d}\tau.
\end{eqnarray}
It follows that
\begin{equation}
R_{\mbox{\tiny T}}(t,s)\dfn R_{\mbox{\tiny
s}}(t,s)+R_{\mbox{\tiny
m}}(t,s)+R_n(t,s)=\overline{g^2}\cdot\int_0^T\lambda(\tau-
\theta)h(t-\tau)h(s-\tau)\mbox{d}\tau+\frac{N_0}{2}\delta(t-s).
\end{equation}
Thus,
\begin{eqnarray}
\bE\{U_n^2\}&=&\int_0^T\int_0^T\mbox{d}t\mbox{d}sR_{\mbox{\tiny
T}}(t,s)\dot{w}(t-\theta)\dot{w}(s-\theta)\nonumber\\
&=&\frac{N_0}{2}\int_0^T\dot{w}^2(t-\theta)\mbox{d}t+\overline{g^2}
\int_0^T\int_0^T\mbox{d}t\mbox{d}s\dot{w}(t-\theta)\dot{w}(s-\theta)\int_0^T
\lambda(\tau-\theta)h(t-\tau)h(s-\tau)\mbox{d}\tau\nonumber\\
&=&\frac{N_0}{2}\int_0^T\dot{w}^2(t-\theta)\mbox{d}t+\overline{g^2}
\int_0^T\mbox{d}\tau\lambda(\tau-\theta)\left[\int_0^T\mbox{d}t\dot{w}(t-\theta)h(t-\tau)\right]
\left[\int_0^T\mbox{d}s\dot{w}(s-\theta)h(s-\tau)\right]\nonumber\\
&\approx&\frac{N_0}{2}\int_0^T\dot{w}^2(t-\theta)\mbox{d}t+\overline{g^2}q_{\mbox{\tiny
e}}^2\int_0^T\lambda(t)\dot{w}^2(t)\mbox{d}t\nonumber\\
&=&\int_0^T\left[\frac{N_0}{2}+\overline{g^2}q_{\mbox{\tiny
e}}^2\lambda(t)\right]\dot{w}^2(t)\mbox{d}t,
\end{eqnarray}
which yields
\begin{equation}
\bE\{(\hat{\theta}-\theta)^2\}\approx
\frac{\int_0^T\left[\frac{N_0}{2}+\overline{g^2}q_{\mbox{\tiny
e}}^2\lambda(t)\right]\dot{w}^2(t)\mbox{d}t}{\overline{g}^2q_{\mbox{\tiny
e}}^2\left[\int_0^T\lambda(t)\ddot{w}(t)\mbox{d}t\right]^2}.
\end{equation}
It is desired to find the optimal waveform $\{w(t)\}$.
First, observe that the MSE is invariant to scaling of $\{w(t)\}$, so the
problem is equivalent to maximizing the absolute value of
\begin{equation}
\int_0^T\lambda(t)\ddot{w}(t)\mbox{d}t
\end{equation}
subject to a given value of
\begin{equation}
\label{cons}
\int_0^T\left[\frac{N_0}{2}+\overline{g^2}q_{\mbox{\tiny
e}}^2\lambda(t)\right]\dot{w}^2(t)\mbox{d}t.
\end{equation}
Let $\omega(t)=\dot{w}(t)$ and 
\begin{equation}
v(t)=\omega(t)\cdot\sqrt{N_0/2+\overline{g^2}q_{\mbox{\tiny
e}}^2\lambda(t)}.
\end{equation}
Then, the problem is equivalent to that of
maximizing the absolute value of 
\begin{equation}
\int_0^T\lambda(t)\cdot\left[\frac{v(t)}{\sqrt{N_0/2+\overline{g^2}q_{\mbox{\tiny
e}}^2\lambda(t)}}\right]'\mbox{d}t,
\end{equation}
or equivalently, 
\begin{equation}
\lambda(t)\cdot\frac{v(t)}{\sqrt{N_0/2+\overline{g^2}q_{\mbox{\tiny
e}}^2\lambda(t)}}\Bigg|_0^T-
\int_0^T\dot{\lambda}(t)\cdot\frac{v(t)}{\sqrt{N_0/2+\overline{g^2}q_{\mbox{\tiny
e}}^2\lambda(t)}}\mbox{d}t,
\end{equation}
subject to a given value of (\ref{cons}).
Now, assuming that 
$\lambda(0)=\lambda(T)=0$, the contribution of the
first term in the last expression vanishes, and so,
we wish to maximize the absolute value of
\begin{equation}
\int_0^T\dot{\lambda}(t)\cdot\frac{v(t)}{\sqrt{N_0/2+\overline{g^2}q_{\mbox{\tiny
e}}^2\lambda(t)}}\mbox{d}t
\end{equation}
for a given energy of $\{v(t)\}$. The maximum is achieved for
\begin{equation}
v(t)=\frac{\dot{\lambda}(t)}{\sqrt{N_0/2+\overline{g^2}q_{\mbox{\tiny
e}}^2\lambda(t)}},
\end{equation}
which yields
\begin{equation}
\omega(t)=\frac{v(t)}{\sqrt{N_0/2+\overline{g^2}q_{\mbox{\tiny
e}}^2\lambda(t)}}=\frac{\dot{\lambda}(t)}{N_0/2+\overline{g^2}q_{\mbox{\tiny
e}}^2\lambda(t)},
\end{equation}
and so, finally,
\begin{equation}
w_*(t)=\int_0^t\frac{\dot{\lambda}(s)\mbox{d}s}{N_0/2+\overline{g^2}q_{\mbox{\tiny
e}}^2\lambda(s)}=\frac{1}{\overline{g^2}q_{\mbox{\tiny e}}^2}
\ln\left[\frac{N_0}{2}+\overline{g^2}q_{\mbox{\tiny e}}^2\lambda(t)\right]+c,
\end{equation}
where $c$ is an integration constant (not to be confused with the earlier
defined constant, $c$).
Equivalently, after a simple manipulation 
of the integration constant, $w_*(t)$ can be presented as
\begin{equation}
w_*(t)\propto\ln\left[1+\frac{2\overline{g^2}q_{\mbox{\tiny
e}}^2}{N_0}\cdot\lambda(t)\right].
\end{equation}
As can be seen, for very large $N_0$, $w(t)$ is approximately proportional to $\lambda(t)$, as expected.
For very small $N_0$, $w(t)$ is proportional to
$\ln\lambda(t)$, which is also expected. Note that if $\lambda(t)$ includes a
dark--current component, $\lambda_{\mbox{\tiny d}}$, then it simply adds to the spectral
density of the thermal noise, $N_0/2$. In other words, there is no distinction
here between the dark--current shot noise and the Gaussian thermal noise.
This should not be surprising, because when calculating the MSE, only second
moments are relevant, and not finer details of the distributions of the
various kinds of noise.

Let us compare the MSE of the optimal correlator $w_*$ to that of the ordinary
matched correlator, $w(t)=\lambda(t)$. Denote
$\lambda_0=2\overline{g^2}q_{\mbox{\tiny e}}^2/N_0$. Then, the MSE of the latter
is given by
\begin{equation}
\mbox{MSE}
=\frac{1}{\lambda_0}\cdot\frac{\int_0^T[1+\lambda(t)/\lambda_0]\dot{\lambda}^2(t)\mbox{d}t}
{\left[\int_0^T\dot{\lambda}^2(t)\mbox{d}t\right]^2},
\end{equation}
whereas the former is given by
\begin{equation}
\mbox{MSE}_{\mbox{\tiny opt}}
=\frac{1}{\lambda_0}\cdot\frac{1}{\int_0^T\frac{\dot{\lambda}^2(t)\mbox{d}t}{1+\lambda(t)/\lambda_0}}.
\end{equation}
It is easy to see that $\mbox{MSE}_{\mbox{\tiny opt}}$ is indeed smaller than
$\mbox{MSE}$, due to the Schwartz--Cauchy inequality, as
\begin{eqnarray}
\left[\int_0^T\dot{\lambda}^2(t)\mbox{d}t\right]^2&=&
\left[\int_0^T\frac{\dot{\lambda}(t)}{\sqrt{1+\lambda(t)/\lambda_0}}\cdot
\dot{\lambda}(t)\sqrt{1+\frac{\lambda(t)}{\lambda_0}}\mbox{d}t\right]^2\nonumber\\
&\le&\left[\int_0^T\frac{\dot{\lambda}^2(t)}{1+\lambda(t)/\lambda_0}\mbox{d}t\right]\cdot
\left[\int_0^T\dot{\lambda}^2(t)\left(1+\frac{\lambda(t)}{\lambda_0}\right)\mbox{d}t\right].
\end{eqnarray}
The difference becomes smaller when
$\dot{\lambda}(t)\sqrt{1+\lambda(t)/\lambda_0}$ becomes
closer to be proportional to
$\dot{\lambda}(t)/\sqrt{1+\lambda(t)/\lambda_0}$, namely, when
$\lambda(t)\ll \lambda_0$, as expected.

This high--SNR analysis can be extended to handle non--white noise and
even non--stationary thermal noise of auto-correlation function
$R_n(s,t)=\bE\{n(s)n(t)\}$. The only difference is that now the numerator of
(\ref{approxerror}) will be replaced by
\begin{equation}
\int_0^T\dot{w}(s)\dot{w}(t)[R_n(s,t)+\overline{g^2}q_{\mbox{\tiny
e}}^2\sqrt{\lambda(s)\lambda(t)}\delta(t-s)]\mbox{d}s\mbox{d}t
\dfn\int_0^T\dot{w}(s)\dot{w}(t)R_0(s,t)\mbox{d}s\mbox{d}t.
\end{equation}
The resulting optimum correlator would then be given by
\begin{equation}
w_*(t)=\int_0^t\mbox{d}\tau\int_0^T\mbox{d}s\dot{\lambda}(s)R_0^{-1}(s,\tau),
\end{equation}
where $R_0^{-1}(s,\tau)$ is the formal inverse 
of the kernel $R_0(s,t)$ (provided that it exists), namely, the
kernel that satisfies
\begin{equation}
\int_0^TR_0(s,t_1)R_0^{-1}(s,t_2)\mbox{d}s=\delta(t_1-t_2).
\end{equation}

The other side of the coin of high--SNR estimation performance is the
probability of anomaly (see, e.g., \cite[Chap. 8]{WJ65}). 
The common practice in assessing this probability is
to divide the interval $[0,T]$ into small bins whose sizes are about the pulse
width and to apply a union bound on the 
(essentially identical) probabilities of the pairwise events that the maximum
$Q(\theta)$ in each bin exceeds the one that contains the correct delay.
Since the number of bins grows only linearly in $T$, the exponent of this
probability is the same as that of each individual bin. The analysis of each
pairwise event can be carried out very similarly to the earlier analysis of
the FA probability (just with a few minor twists), whose dependence on the mismatched correlator $w$ is
different that of the above derived approximated MSE. 
Thus, the signal design should seek a compromise that trades off
the weak--noise MSE with the probability of anomaly. In the high--SNR regime, a
logarithmic correlator is good both for the MSE and for keeping the anomaly
probability small.

\section*{Acknowledgment}

Interesting discussions with Moshe Nazarathy, at the early stages of this work, are
acknowledged with thanks.

%\section*{Appendix}
%\renewcommand{\theequation}{A.\arabic{equation}}
%    \setcounter{equation}{0}

\end{document}